\documentclass[two column, apl]{revtex4}
\usepackage{bm}
\usepackage{graphics}
\usepackage{bm}
\usepackage{graphics}
\usepackage{graphicx}
\usepackage{amsfonts}
\usepackage{amsmath}
\usepackage{amssymb}
\usepackage{graphicx}
\usepackage{color}
\usepackage{fancyhdr}
\usepackage{soul}

\begin{document}

\title{Nano-Diamond Thin Film Field Emitter Cartridge for Miniature High Gradient Radiofrequency $X$-band Electron Injector}

\author{Jiaqi Qiu$^1$}
\author{Stanislav S. Baturin$^2$}
\author{Kiran K. Kovi$^1$}
\author{Oksana Chubenko$^1${\footnote{On leave from the George Washington
University}}}
\author{Gongxiaohui Chen$^1$\footnote{On leave from Illinois Institute of Technology}}
\author{Richard Konecny$^1$}
\author{Sergey Antipov$^1$}
\author{Chunguang Jing$^1$}
\author{Anirudha V. Sumant$^3$}
\author{Sergey V. Baryshev$^1$\footnote{Current address: The Department of Electrical and Computer Engineering, Michigan State University, 428 S. Shaw Lane, East Lansing, MI 48824, USA}\footnote{sergey.v.baryshev@gmail.com}\vspace{0.3cm}}

\affiliation{$^1$Euclid TechLabs, 365 Remington Blvd.,
Bolingbrook, IL 60440, USA
\\
$^2$PSD Enrico Fermi Institute, The University of Chicago, 5640 S.
Ellis Ave., Chicago, IL 60637, USA
\\
$^2$Center for Nanoscale Materials, 9700 S. Cass Ave., Argonne, IL
60439, USA}

\begin{abstract}

\noindent The complete design, fabrication, and performance
evaluation of a compact, single cell, $X$-band ($\sim$9 GHz)
electron injector based on a field emission cathode (FEC) are
presented. A pulsed electron beam is generated by a 10's of kW
radiofrequency (RF) magnetron signal from a plug-in thin film
nitrogen-incorporated ultrananocrystalline diamond (N)UNCD FEC
cartridge. Testing of the $X$-band injector with the (N)UNCD FEC was
conducted in a beamline equipped with a solenoid, Faraday cup and
imaging screen. The results show that typically the (N)UNCD FEC
cartridge produces $\gtrsim$1 mA/cm$^2$ at a surface electric
field of 28 MV/m. The diameter of the output beam generated from
the 4.4 mm diameter (N)UNCD cartridge can be as small as 1 mm. In
terms of its practical applications, the demonstrated $X$-band
electron injector with the (N)UNCD plug-in FEC can serve as a
source for X-ray generation, materials processing, travelling-wave
tubes (including GHz and THz backward wave oscillators), or can be
used to drive slow-wave accelerating structures. The results presented
also suggest that this field emitter technology based on
planar (N)UNCD thin films, which are simply grown on the surface
of optically polished stainless steel, can enable a vast number of
device configurations that are efficient, flexible in design, and
can be packaged with ease.

\end{abstract}

\maketitle \pagenumbering{gobble} Radiofrequency (RF) electron
guns that drive industrial, medical and scientific accelerators
typically use electric fields in excess of 10-20 MV/m, applied to
the cathode surface. This is to mitigate space-charge effects for
efficient beam transport and downstream staged acceleration, or to
produce high energy electrons for efficient interaction with
slow-wave structures in vacuum electronics devices. A number of
accelerator applications require high brightness, highly coherent
beams (free electron lasers) making use of photocathodes, or high
current beams (synchrotrons, industrial electron beam processing
and cargo inspection facilities) making use of thermionic
emitters. Field emission cathodes (FECs) could occupy a niche in
which moderate beam currents are needed, and/or in which
simplicity and compactness of the electron injection system (as
compared to laser-photocathode concept) and/or fast ON/OFF
switching (as compared to slow thermionic sources) is a must.
Therefore, some efficient, simple and scalable in design and
fabrication electron field emission source would be needed for
high gradient injection, and the search is ongoing [1-4].

Compared to thermionic and photoemission cathodes, FECs have kept
a relatively low application profile. Development of a vast number
of FECs based on metallic and semiconductor whiskers/wires [5],
pillars [6], tips [7] etc. did lead to a limited number of high
power [8, 9] or high speed devices [10, 11] and other applications
of field emitters [11]. Even so, the major progress to date in
device fabrication and commercialization was achieved using only
one field emitter platform, the proprietary molybdenum Spindt
cathodes developed and manufactured at Stanford Research Institute
[11, 13]. The standard problems remain the same: geometrical
inhomogeneity of emitters in the array and fabrication complexity
that make scalable FEC production challenging. Carbon-derived
materials such as carbon nanotubes (CNTs), on the other hand, have
gained greater attention as field emission sources. They are
simple to synthesize on various substrates of various form factors
and their synthesis can be scaled. Although CNTs have proven to be
remarkable FECs [14], the applied macroscopic electric fields have
to be maintained in the range below 10 MV/m because CNTs (like any
other sharp emitters) have an extremely high aspect ratio leading
to a large geometrical enhancement factor $\beta$, on the order of
1,000. High aspect ratio structures with $\beta$$\sim$1,000 can be
impaired or disrupted when placed in an external electric field
larger than 10 MV/m, because the product of the external
macroscopic electric field multiplied by $\beta$, called the local
electric field, sets the ultimate field limit a solid state
material can physically sustain; this ultimate field is $\sim$10
V/nm [15, 16]. Of course, the field emitter failure is also
heavily dependent on the local current density through the
emitter, but the rough magnitude of the external macroscopic
electric field applied to a field emitter can frequently be
estimated as less than or on the order of $\frac{10 V/nm}{\beta}$.
This limiting factor precludes CNT FECs from use as electron
sources in high gradient electron injectors.

Previously, it was established that some other carbonic materials,
such as amorphous carbons and polycrystalline diamond, e.g., in
the form of nano-crystalline and ultra-nano-crystalline diamond
can be effective field emitters in the planar thin film
configuration [17, 18]. More specifically, nitrogen-incorporated
ultrananocrystalline diamond ((N)UNCD) field emitters have long
life times [19] and low turn-on macroscopic fields $\lesssim$10
MV/m [18, 20]. Thin film (N)UNCD FEC can be operated at electric
fields $\gtrsim$10 MV/m, i.e. in the field range that is
complementary to that of CNTs, and make (N)UNCD attractive for
high gradient electron injector technology. Following our previous
proof-of-concept experiments with $L$-band 1.3 GHz MW-class
electron injectors [3], we now present results on the performance
of a (N)UNCD planar field emitter installed in a custom, miniature
$X$-band 9.17 GHz, kW-class injector driven by a conventional RF
magnetron delivering $<$60 kW of power which produced up to 28
MV/m surface gradient. The reason for the choice of this
application case is two-fold. First to demonstrate the scalability
of the injection concept that relies on the (N)UNCD FEC by making
a complete and properly working device miniaturized 10-fold as
compared to $L$-band type injectors [3]. Second, to demonstrate
that (N)UNCD plug-in FEC is a technology that can enable
fabrication, commissioning and operation of most miniature
electron guns in the RF electron injector family (e.g. $X$-band
spans 8-12 GHz, while the RF gun size is inversely proportional to
the operating frequency $f$). In this case, size constraints on
the use of laser injection or high temperature heater installation
can be fully avoided. The injector presented here has a
conventional single-cell quarter-wave design. It combines a copper
body that stores the energy of the RF field with stainless steel
flanges. Circular standard 1.33" conflat type flanges at the front
and the back of the injector provide for vacuum beamline
connections and cathode plug insertion. A CERN type rectangular
flange provides for vacuum/RF connection. The copper and stainless
steel parts were brazed together. Fig.1a illustrates the final
injector assembly together with the cathode plug, and the
simulated electric field distribution. Fig.1b shows the simulated
versus measured $S_{11}$-parameter (input port reflection). The
$S_{11}$ measurement was made in a blanked cold test, i.e. a cold
test of the empty cavity with no cathode installed. Its
fundamental frequency was found to be 9.17 GHz.

\begin{figure}[!t]
\includegraphics[height=3.4cm]{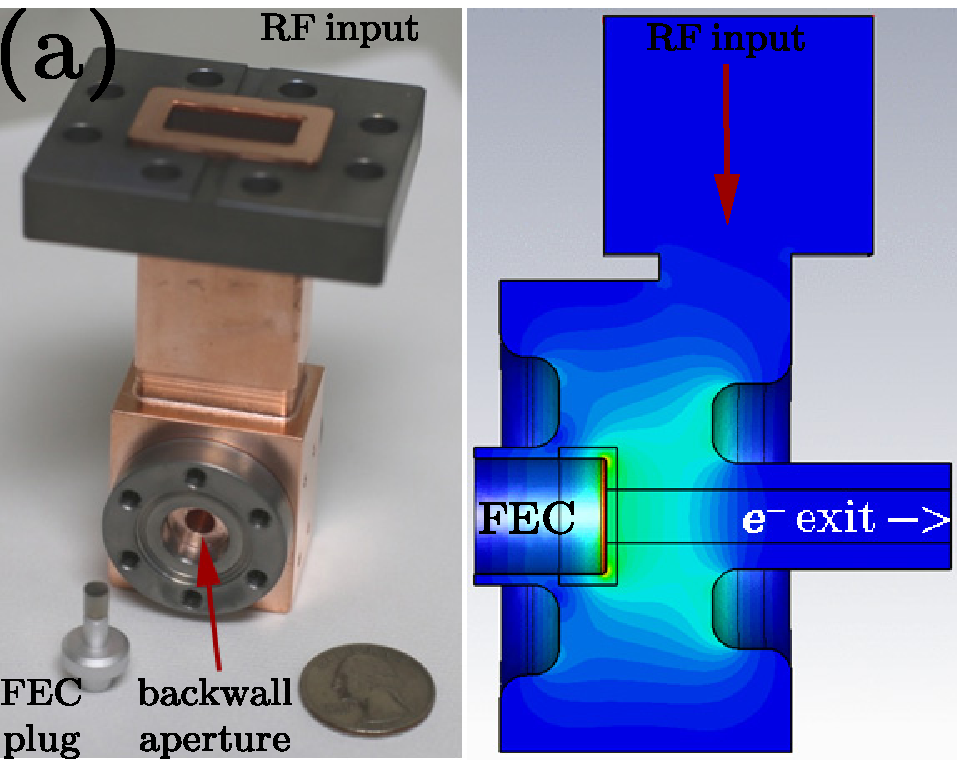} \includegraphics[height=3.4cm]{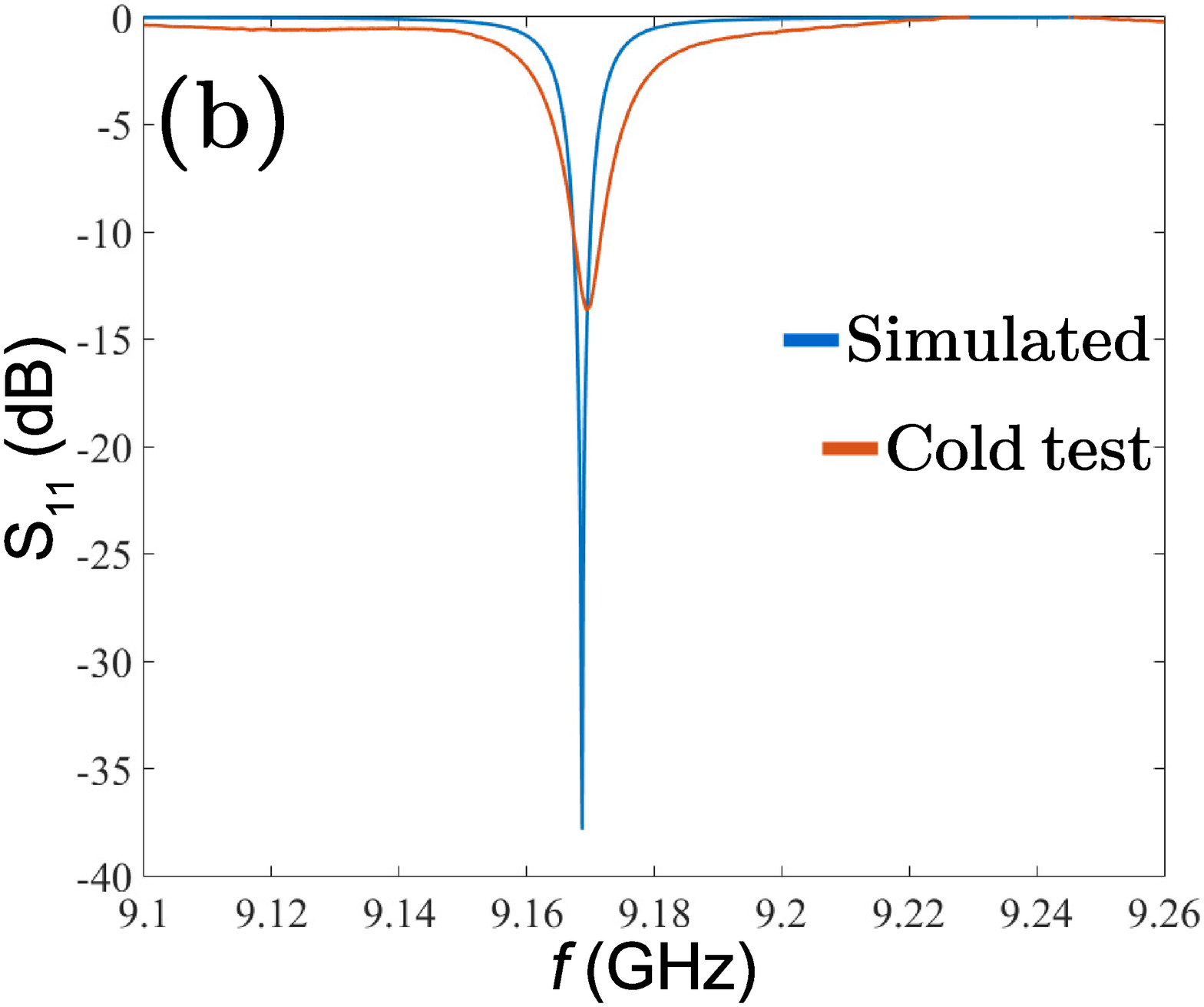}

\caption{(a) The machined and brazed injector in its final form
photographed together with the cathode plug adjacent to the back
wall aperture where it will be installed, and its RF electric
field distribution computed using CST Microwave Studio. (b) The
measured (red trace) and simulated (blue trace) frequency
dependence of the input power reflection parameter
($S_{11}$).\vspace{-0.5cm}}
\end{figure}

The full cathode plug assembly is shown in Fig.2. It consists of a
base with a venting hole on the back. The cartridge made of
stainless steel (SS) is attached to the base with a set screw. At
the joint location, there is a groove that holds a spring, the
standard technique for enabling good RF contact between the
cathode plug and the body of the injector. The SS cartridge was
mirror polished and coated with Mo. Subsequently the (N)UNCD thin film
was deposited using the standard microwave-plasma assisted CVD
process [3].

\begin{figure}[!t]
\includegraphics[width=6cm]{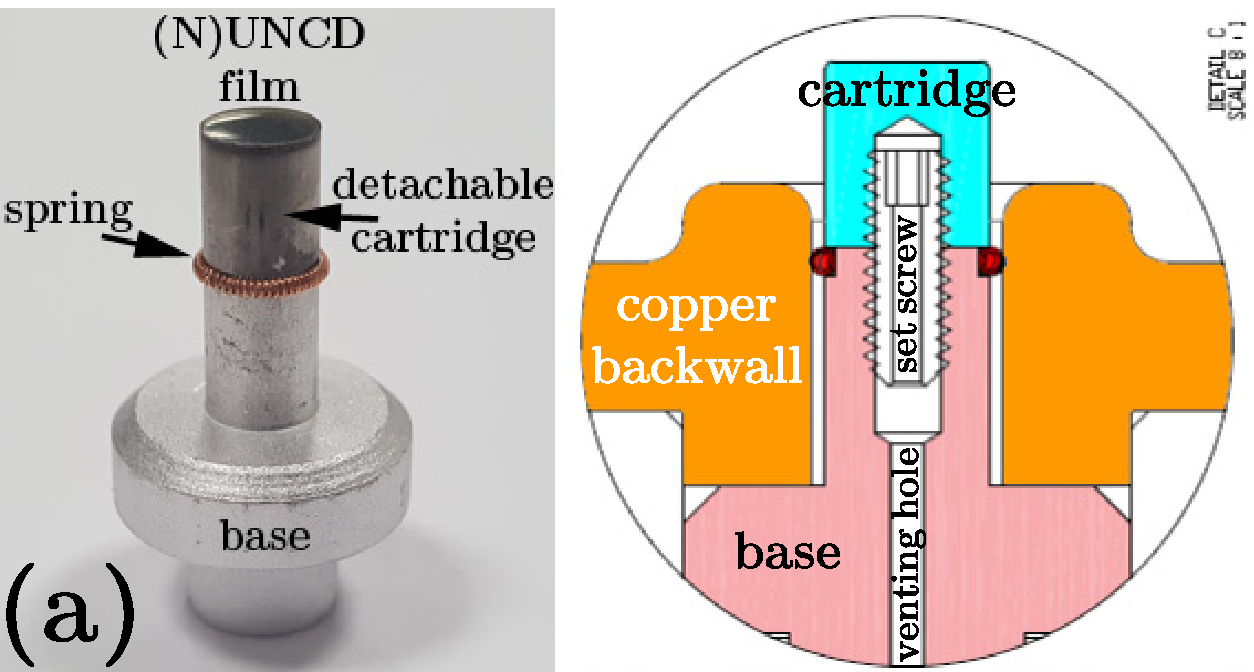}

\includegraphics[width=6cm]{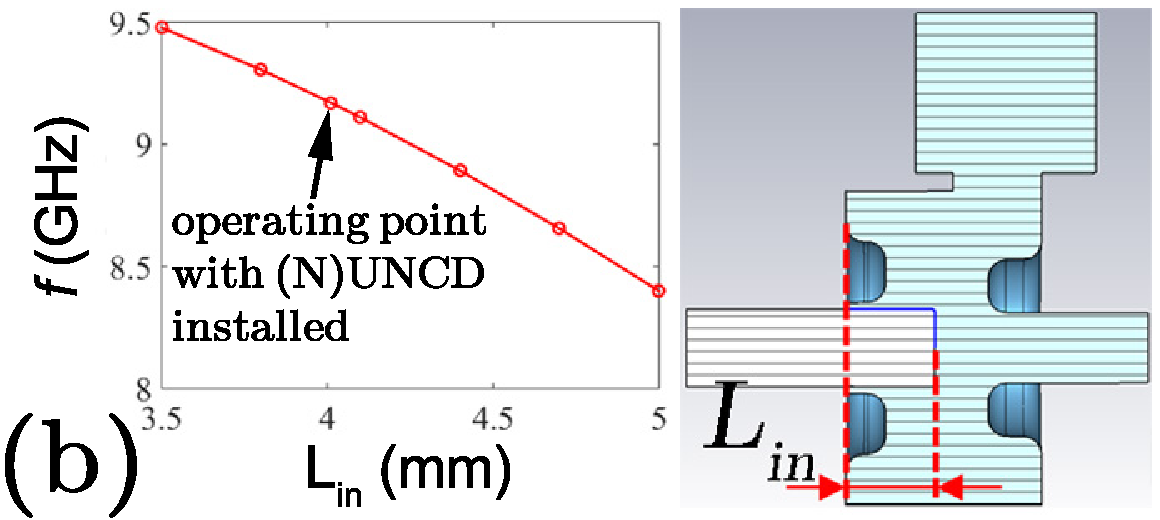}

\caption{(a) A photograph of the actual cathode plug with the
(N)UNCD/Mo/SS cartridge on top, and its mechanical design when
installed in the back wall aperture of the injector. (b) The
dependence of the resonance frequency of the cavity on the length
of the cathode extending inside the cavity. The right panel
represents the cross section view of the gun and demonstrates how
the parameter $L_{in}$ is defined.\vspace{-0.5cm}}
\end{figure}

Upon installation, the length of the cartridge inside the cavity
determines the operating frequency by changing the effective size
of the cavity volume seen by the RF drive signal. The total length
of the cathode plug tip with the cartridge atop is precisely
machined. From the known geometry, simulations and cold testing,
we can relate the final frequency as measured with a network analyzer
to the length of the cartridge inside the cavity, which cannot be
directly measured. This is an important correlation because it
affects the distance between the (N)UNCD field emitter and the
exit aperture plane and thus the ultimate energy gain of the
output electrons (product of on-axis electric field and distance).
In Fig.2b, the resonance frequency is plotted against the
cartridge length inside the copper injector cavity, the $L_{in}$
parameter. The way the length $L_{in}$ is defined is depicted on
the right panel of Fig.2b. The final cathode plug design, when
fully assembled and installed with the 1.33" flange tightened,
leads to $L_{in}$=4 mm, thus fixing the operating frequency at
9.17 GHz. The full set of the key parameters that describe the
performance of the injector are summarized in Table 1.

\

\noindent \textbf{Table 1.} The simulated and measured fundamental
parameters of the fabricated cavity with the (N)UNCD/Mo/SS
cartridge installed: $Q_0$   is the cavity quality factor,
$Q_{external}$  is the
quality factor of the whole injector assembly that includes RF
window and waveguide losses.

\begin{center}
\begin{tabular}{ c|c|c }
   & simulation & cold test \\
 \hline
 $f_0$ (GHz) & 9.1697 & 9.1694 \\
 \hline
 $Q_0$ & 2286 & 800 \\
 \hline
 $Q_{external}$ & 2285 & 1218 \\
 \hline
 $b$=$Q_0$/$Q_{external}$ & 1.00 & 0.66 \\
 \hline
 $E^2_{cath}$/$P_{loss}$ & 4.3$\times10^{10}$ & n/a \\
\end{tabular}
\end{center}

Fig.3 shows the full beamline layout. The entire beamline assembly
is located on top of an oil-free hybrid diaphragm/turbo pump
station. The working pressure in the beamline is $\sim
1\times10^{-8}$ Torr. The beamline consists of:

\noindent 1.  The cavity connected to the magnetron through an
$X$-band RF window, which is an RF transparent septum between the
vacuum volume of the cavity and ambient pressure waveguide from
the magnetron;

\noindent 2.  The solenoid installed outside the vacuum pipe directly
at the cavity exit to focus the exiting beam;

\noindent 3.  Specialty \emph{in vacuum} YAG:Ce screen, or a
Faraday cup with a telescopic electrode which can collect
electrons either at the YAG:Ce screen position or directly at the
injector exit aperture. Only the YAG:Ce screen or the Faraday
cup can be installed at a given time at the end of the
beamline, with either detector serving also as a beam dump;

\noindent 4.  Steering magnet at the exit plane of the solenoid to
steer the beam in the transverse plane across the YAG screen to
measure the beam energy.

\begin{figure}[!t]
\includegraphics[width=7cm]{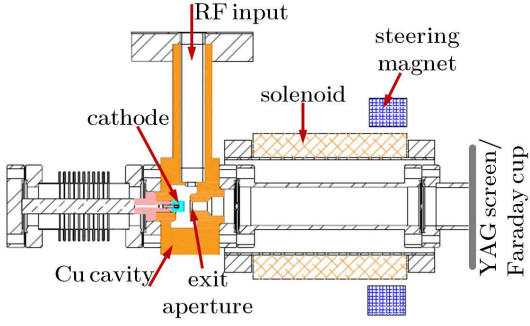}

\caption{Cross-section of the beamline that depicts all the key
components.}\vspace{-0.5cm}
\end{figure}

The magnetron produces a 1 $\mu$s macroscopic pulse (Fig.4a)
consisting of 9.17 GHz oscillations at a power of up to 250 kW. The RF power is split with a 6 dB splitter, so that the maximum
power available in our experiment is 57 kW. Using the values of the injector
parameters listed in Table 1, the input magnetron power can be
converted into the electric field on the cathode surface
$E_{cath}(z=0)$ as

\[
E^{exp}_{cath}=\sqrt{\bigg(\frac{E^2_{cath}}{P_{loss}}\bigg)^{sim}\cdot
P_{in}\cdot\frac{4b}{(1+b)^2}\cdot}\sqrt{\frac{Q^{exp}_0}{Q^{sim}_0}}
 \text{, (1)}
\]
\noindent where $exp$ stands for experiment, $sim$ stands for
simulation, and $in$ stands for input.

The RF pulse in Fig.4a demonstrates the typical behavior: two
sharp RF reflection features at the beginning and the end of the
pulse related to filling with and emptying the cavity of RF
energy. Fig.4b shows temporal profile of a Faraday cup signal. A
standard $RC$ measurement circuit with a time constant of 1 ms was
used. The actual full charge (measured in C) per 1 $\mu$s RF
macroscopic pulse extracted from the cathode and transported
through the exit aperture is given by
\[
\frac{\int V(t)dt}{R} \text{, (2)}
\]
\noindent where $V(t)$ is the scope trace of voltage vs time, and
$R$ is the resistor value equal to 1 M$\Omega$ (with capacitor
value of 1 nF).

\begin{figure}[!t]
\includegraphics[width=5cm]{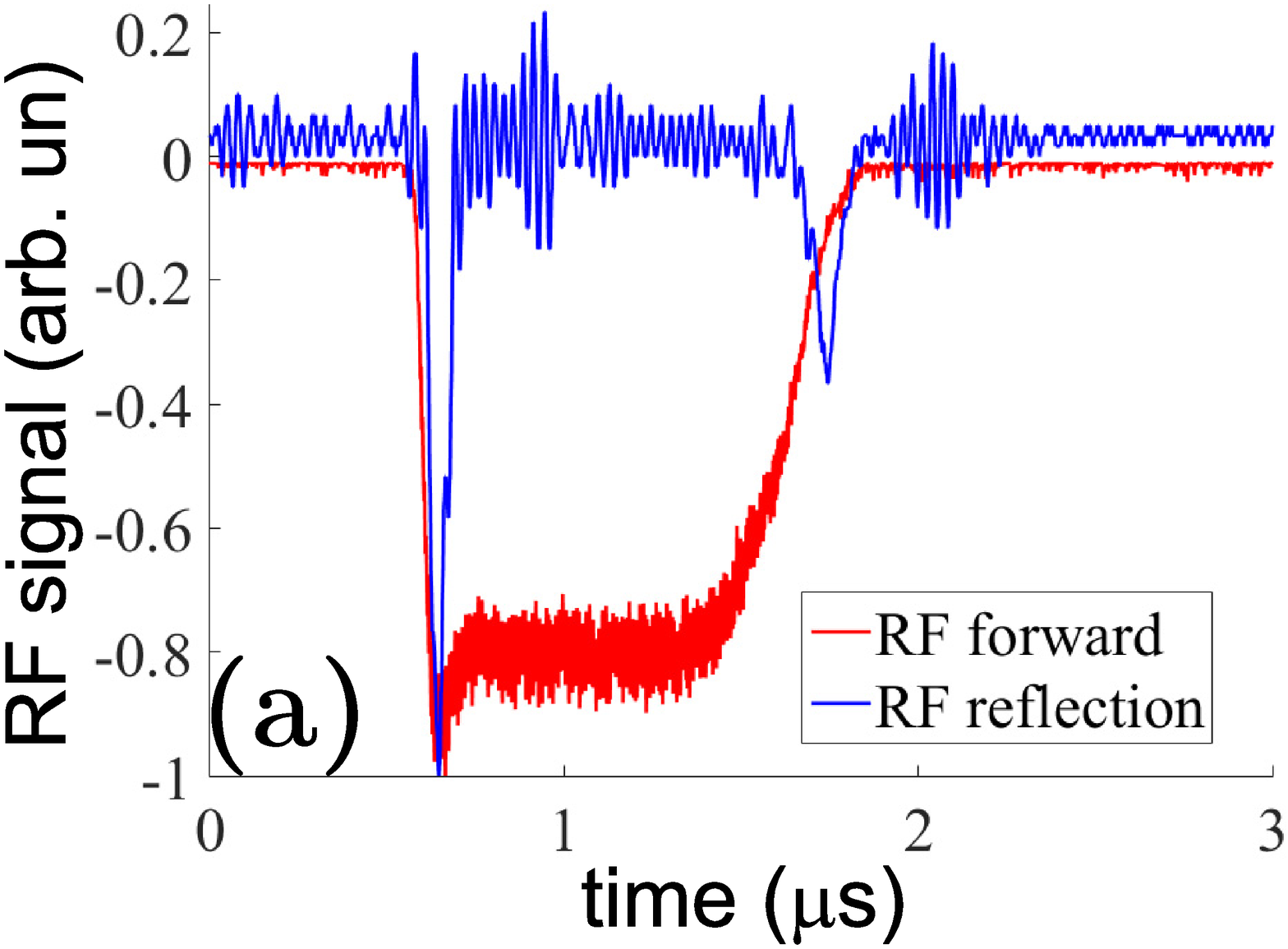}

\includegraphics[width=5cm]{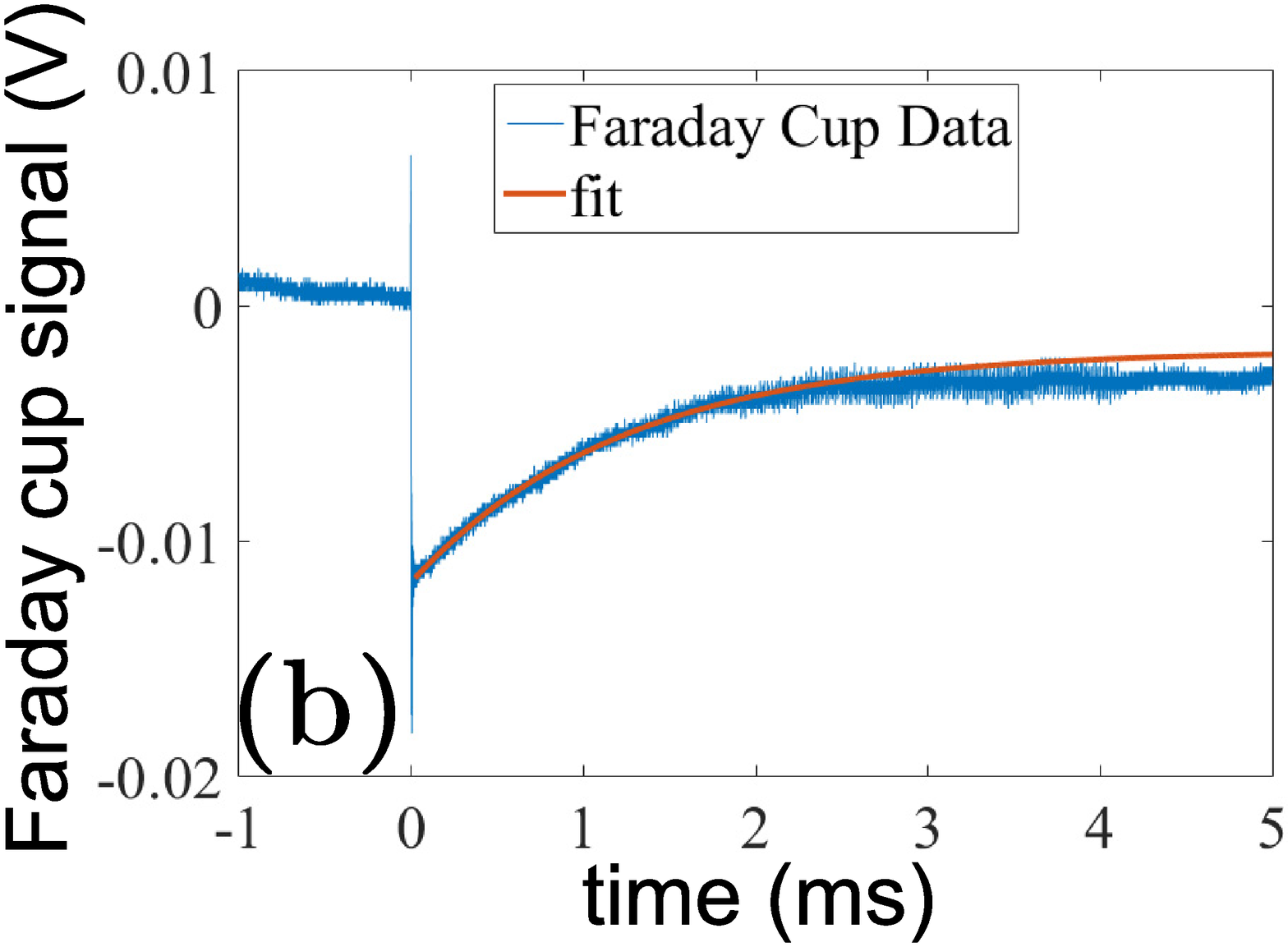}

\includegraphics[width=5cm]{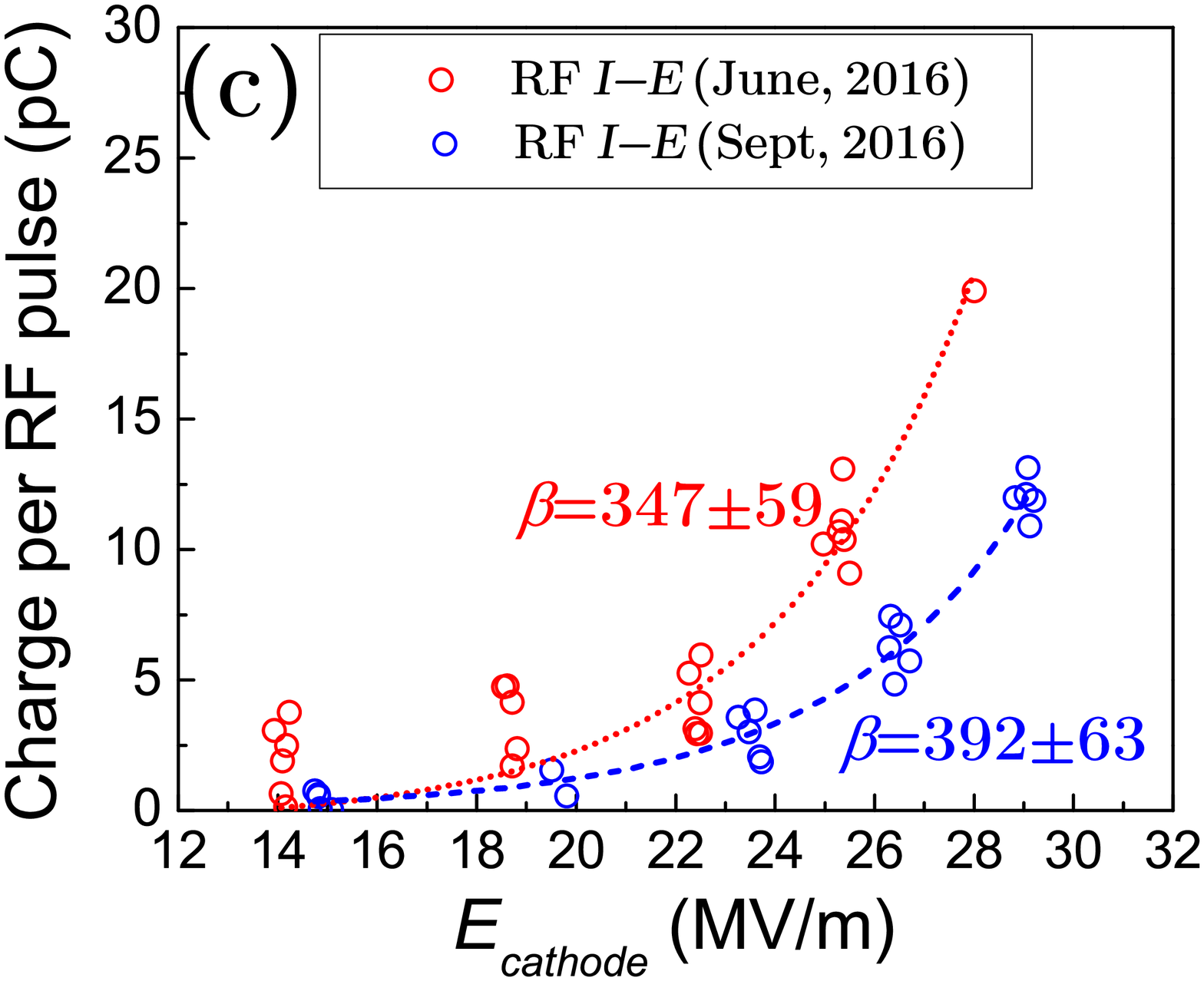}

\caption{(a) Scope traces of the RF pulse in terms of absorbed and
reflected RF power. (b) Scope trace of a Faraday cup signal. (c)
The $Q$-$E$ characteristics calculated from the recorded scope
traces of the RF pulse and the $RC$-signal of the Faraday cup,
namely, RF pulse charge versus $E_{cath}$ in linear coordinates
(dotted and dashed lines are only to guide the
eye).\vspace{-0.5cm}}
\end{figure}

Finally, using Eqs. (1) and (2) we obtain the dependence of the
charge per RF pulse versus $E_{cath}$, as shown in Fig.4c. The
cathode was tested over a three month period with the cathode
performing under high electric field for a total of two weeks continuously at
a 10 Hz repetition rate. Data were taken at 1 Hz.
Fig.4c represents data from the first and the last runs. As can be
seen, the cathode was conditioned to up to 28 MV/m, and a charge
per RF pulse of 20 pC was achieved which translates into a current
of $\sim$1 mA/cm$^2$ per RF pulse, if, as before [3], it is
assumed that emission is triggered over $\pm30^\circ$ around the
RF electric field maximum (1/6 the duration of the positive part
of the GHz oscillation) from the entire cartridge area. 20 pC per
RF pulse implies $\sim10^8$ electrons.

Referring to Fig.4c, the Faraday cup telescopic electrode measured
similar values of charge per RF pulse when placed directly at the
exit aperture of the injector or a few centimeters after the
solenoid, with the solenoid current set to 8-10 A corresponding to
about 140 Gauss on the $z$-axis. With the solenoid settings
determined, the Faraday cup was replaced with the \emph{in vacuum}
YAG:Ce screen assembled on the vacuum side of a conflat glass
window with a camera placed behind the window. In this
configuration no imaging distortions were introduced as all
components were placed one after another and centered with respect
to the beam propagation direction ($z$-axis). Fig.5 summarizes the transverse
($x$, $y$) beam profiles that were obtained at different solenoid
settings. At 8.5 A, the beam profile had minimal size with an
intense core of about 1 mm in diameter.

\begin{figure}[!t]
\includegraphics[width=8.5cm]{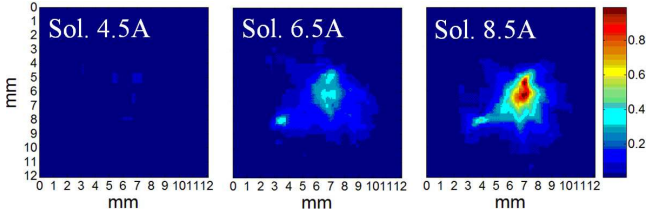}
\caption{YAG:Ce screen images of transverse beam profile at
different solenoid settings. The intensity is normalized in each
panel.\vspace{-0.5cm}}
\end{figure}

\begin{figure}[!t]
\includegraphics[height=3.3cm]{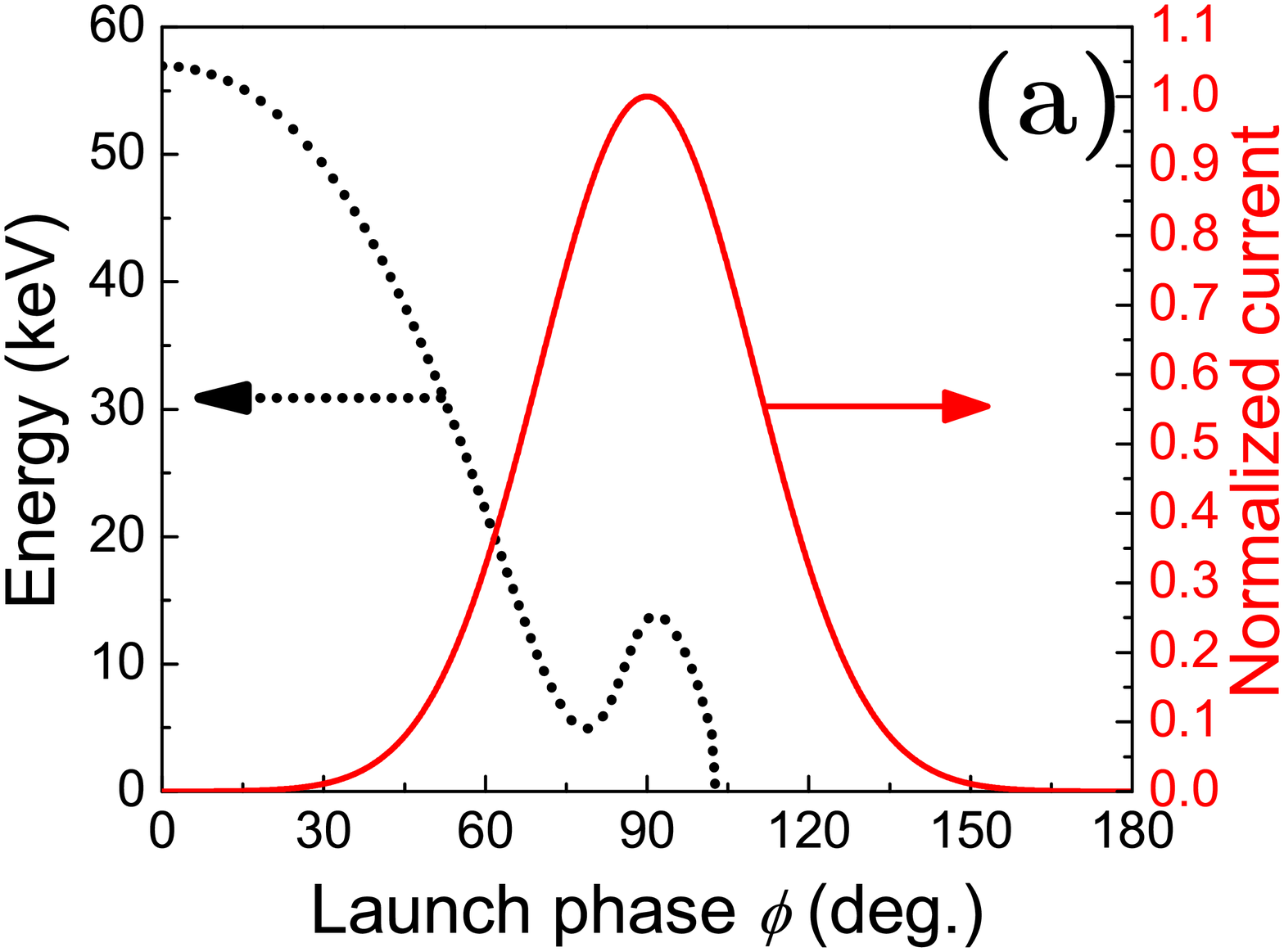} \includegraphics[height=3.3cm]{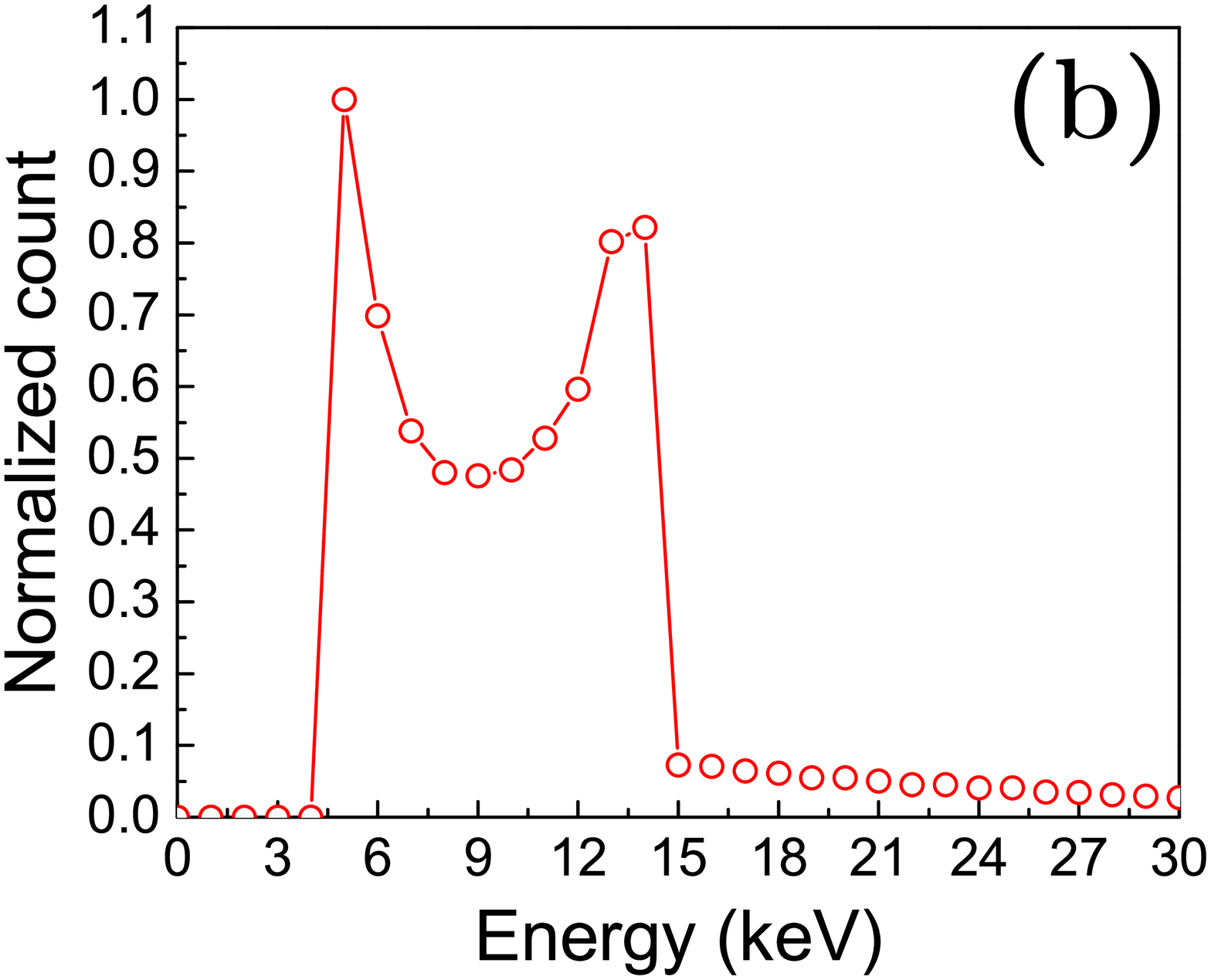}

\includegraphics[height=3.3cm]{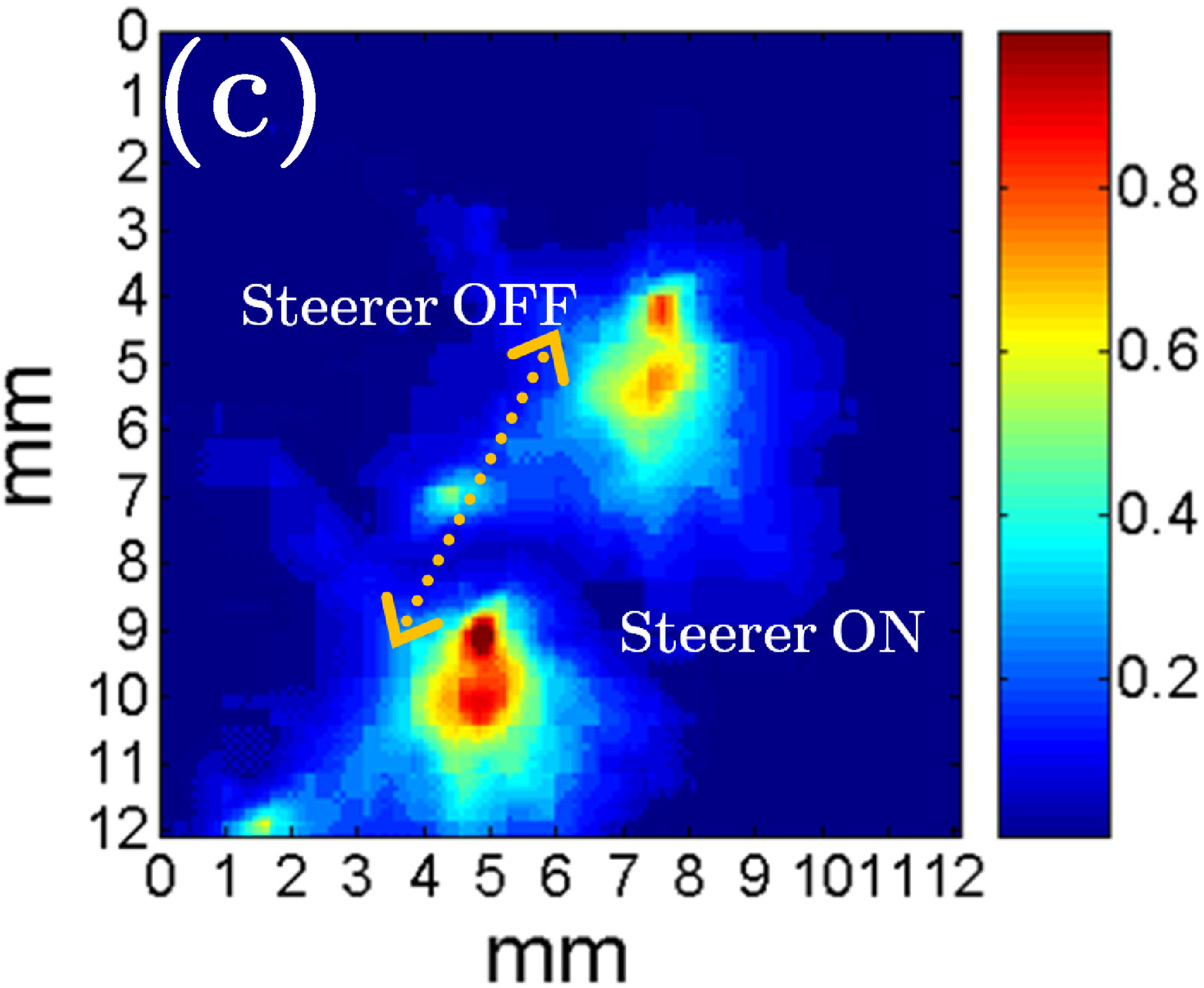} \raisebox{0.2\height}{ \includegraphics[height=2.5cm]{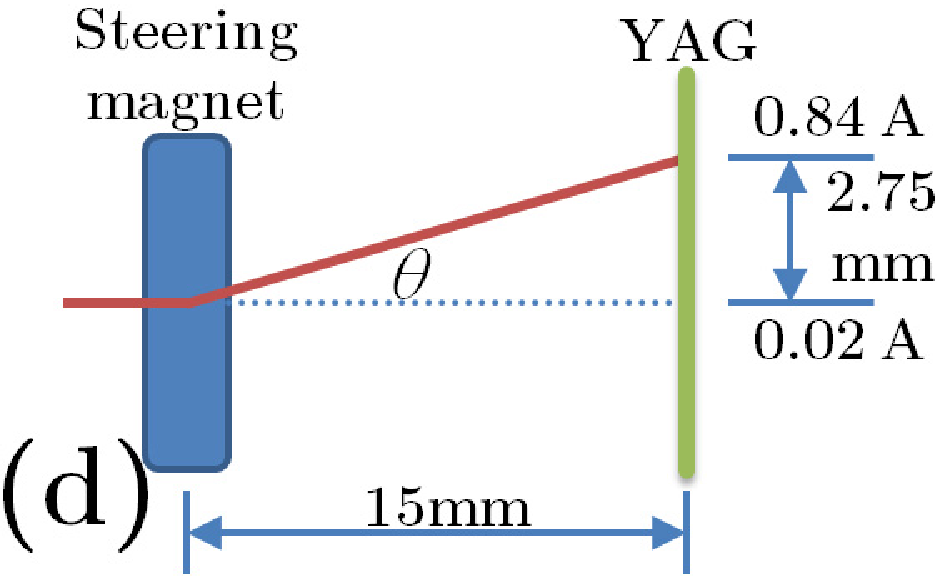}}

\caption{(a) Launch phase dependence of the energy gain of an
electron in the cavity (black dotted line) superimposed on the
current directly calculated from the F-N equation with the
experimentally measured $\beta$ (red solid line). (b) The
resulting electron energy spectrum. (c) YAG:Ce screen images of
two superimposed snapshots with the steering magnet on and off
that enable the beam energy estimation. (d) Diagram used for beam
energy evaluation using Fig.6c and Eq.(3).\vspace{-0.5cm}}
\end{figure}

As before [3], we overlap launch phase dependences of the current
and the energy gain in the gun (Fig.6a) to obtain the resulting
energy spectrum of emitted electrons, as plotted in Fig.6b. The
resulting energy spread is rather wide and spans 5 to 14 keV, as
the cavity is not purposely optimized for better energy spread.
With the solenoid set to 8.5 A, a horizontal scan was performed
with a resulting shift illustrated in Fig.6c; current through the
steering magnet was 0.82 A. Using the measured beam position shift
on the YAG:Ce screen (2.75 mm) and the drift distance from the
steering magnet to the YAG:Ce screen (15 mm), the beam energy
$\varepsilon_k$ was estimated equal to 10 keV (see Fig.6d for
reference). The experimentally measured energy is in excellent
agreement with the 9-10 keV central energy of the spectrum
obtained from the semi-empirical calculations (Fig.6b). The
following formula was used [21]

\[
\beta_r\cdot\varepsilon [keV]=\frac{0.29979}{\theta}\cdot\int Bdl
[G\cdot cm/rad] \text{, (3)}
\]

\noindent where $\varepsilon=\varepsilon_k + \varepsilon_0$, a sum
of the kinetic and rest energy of the electron, is the total
energy; $\beta_r=\frac{v}{c}$ is the ratio of electron velocity
and the speed of light; $\int Bdl$ = 61 G$\cdot$cm was calculated
from the field maps obtained from the steering magnet manufacturer
at a bias current of 0.82 A. Because neither $\beta_r$ nor
$\varepsilon$ are known, Eq.(3) can be solved, e.g., graphically
in order to deduce $\beta_r$ and hence $\varepsilon_k$. Note, the
observed bimodal structure in the launch phase dependence, Fig.6a,
translates into a bimodal structure in the energy spectrum and
into the longitudinal bunch modulation. Due to slight asymmetry in
the injector geometry design and because our imaging system is
time integrating over a span of milliseconds, the longitudinal
bunch modulation projects at different locations on the YAG:Ce
screen (Fig.5 and Fig.6c). This effect does not reflect
non-uniformity of electron emission from the cathode surface.

Built upon our previous results presented for the $L$-band
injector [3], the results presented here suggest that the
technology of plug-in (N)UNCD FEC cartridges developed here
enables simple electron gun devices. A miniature 9.17 GHz injector
device delivering 1 mA/cm$^2$ of current at 10 keV was designed,
fabricated and commissioned. The same approach can be implemented
for a wide range of electron injectors (e.g. $L$-, $S$-, $C$- or
$X$- bands). This means that the proposed concept is essentially
scalable and adaptive. The cartridges are disposable or can be
recycled for (N)UNCD re-deposition. In the earlier work [3], it
was demonstrated that (N)UNCD can survive and perform under fields
as high as 40-65 MV/m. In present work, it is shown that (N)UNCD
field emitters can, having a turn-on field around 5 MV/m, work at
much lower gradients, from 14 to 28 MV/m. These observations let
us conclude that the planar (N)UNCD field emitter can adapt,
without disruption, to the electric field required for a specific
application, starting at $\sim$10 MV/m ($\beta$-factor 300-400)
and higher (at least up to 70 MV/m, $\beta$-factor 100), by
irreversibly reducing its apparent originally high $\beta$-factor
[17, 18], such that the entire operating $I$-$E$ curve is modified
accordingly.

In engineering physics, copper resonant cavity injector
technology is mature. If the field emission current enabled by
(N)UNCD (with no laser or heaters required) is satisfactory for the application, a
half-, single-, or multi-cell injector with RF input ports can be
designed to meet specific requirements on the operating frequency,
power consumption, duty cycle, beam kinetic energy and energy
spread, and transverse phase space.

\

\noindent \textbf{Acknowledgments}

We thank Paul Schoessow for his help with the manuscript. Euclid
was supported by the Office of Nuclear Physics of DOE through a
Small Business Innovative Research grant No. DE SC 0013145.
(N)UNCD synthesis and characterization work at Center for
Nanoscale Materials, Argonne National Laboratory was funded by
Office of Energy Efficiency \& Renewable Energy, Advanced
Manufacturing Office, Department of Energy Technology
Commercialization Fund, under Agreement number 32138, Funding
action number 69530. Use of the Center for Nanoscale Materials, an
Office of Science user facility, was supported by the U.S.
Department of Energy, Office of Science, Office of Basic Energy
Sciences, under Contract No. DE-AC02-06CH11357.

\end{document}